\documentclass[fleqn,10pt]{olplainarticle}
\usepackage{parskip}

\usepackage{enumitem}
\usepackage{graphicx}
\usepackage{listings}
\usepackage{kotex}
\usepackage{amssymb}
\usepackage{amsmath}
\usepackage{tabularx}
\usepackage[normalem]{ulem}

\usepackage{amsmath,amsfonts,bm}

\def\eqref#1{equation~\ref{#1}}
\def\1{\bm{1}}

\DeclareMathAlphabet{\mathsfit}{\encodingdefault}{\sfdefault}{m}{sl}
\SetMathAlphabet{\mathsfit}{bold}{\encodingdefault}{\sfdefault}{bx}{n}

\usepackage{hyperref}
\usepackage{url}

\renewcommand\thesection{\arabic{section}}
\renewcommand\thesubsection{\thesection.\arabic{subsection}}
\renewcommand\thesubsubsection{\thesubsection.\arabic{subsubsection}}

\title{Discovering Models for Experimental Quantum Systems with Agentic AI}

\author[1]{Hannah C. Kleidermacher}
\author[2]{Bennet Windt}
\author[3,4]{Miguel Bello}
\author[5,6*]{James Zou}
\author[1*]{Jelena Vučković}
\affil[1]{Edward L. Ginzton Laboratory, Stanford University, Stanford, CA, USA.}
\affil[2]{Max Planck Institute of Quantum Optics, Garching, Germany.}
\affil[3]{Departamento de Física Teórica de la Materia Condensada, Universidad Autónoma de Madrid, E-28049 Madrid, Spain.}
\affil[4]{Condensed Matter Physics Center (IFIMAC), Universidad Autónoma de Madrid, E-28049 Madrid, Spain.}
\affil[5]{Department of Biomedical Data Science, Stanford University, Stanford, CA, USA.}
\affil[6]{Department of Computer Science, Stanford University, Stanford, CA, USA.}
\affil[*]{\texttt{jamesz@stanford.edu, jela@stanford.edu}}

\begin{abstract}
A central obstacle in realizing color center cavity-QED-based quantum technologies is inhomogeneity, or disorder, among individual emitters. As the number of emitters increases, so does the complexity of the experimental data, making it difficult to quantify the present physical mechanisms of disorder. Here we introduce an artificial intelligence-driven model discovery framework in which a large language model (LLM)-based agent, equipped with tools and a structured workflow, proposes candidate Hamiltonian and Lindbladian models, generates computational tools, and quantitatively evaluates competing physical hypotheses. We apply the workflow to three previously published experimental cavity-QED datasets collected from color centers in diamond and silicon carbide. Across the experiments, the agent successfully identifies physical models for the system and characterizes the disorder. In assessing reproducibility, we find that run-to-run variability is concentrated primarily in the research plan generation step, while the computational analysis itself is more consistent. Our results suggest that structured LLM agent workflows can support model discovery from quantum optical experimental data while simultaneously allowing for human expert supervision and collaboration.
\end{abstract}

\begin{document}
\maketitle

\section{Introduction}
Solid-state quantum systems based on defects in wide-bandgap semiconductors (``color centers'') are among the most promising platforms for quantum sensing, quantum networking, and analogue quantum simulation~\citep{awschalom2018quantum, atature2018material}. In particular, tin-vacancy (SnV$^-$) color centers in diamond have recently emerged as a promising platform following demonstrations of high-fidelity spin control, single-shot electron-spin readout, and quantum photonic interfaces in thin-film diamond~\citep{iwasaki2017tin, thiering2018ab, rosenthal2023microwave, rosenthal2024single, lee2026quantum, codreanu2026above}. Additionally, color centers in silicon carbide (SiC), including silicon-vacancy (V$_{\text{Si}}$) defects in 4H-SiC, offer optically addressable spin qubits within a technologically mature semiconductor platform~\citep{castelletto2020silicon, liu2024silicon, lukin2023two, lukin2025mesoscopic}. These advances establish a strong foundation for scaling quantum technologies.

A central obstacle to scaling quantum hardware is non-uniformity among individual qubits~\citep{de2021materials, janitz2020cavity}. In cavity-QED color-center systems, disorder arises through uncontrolled variation in an emitter's spatial position and cavity coupling strength, inhomogeneous transition frequencies among emitters, and time-dependent spectral diffusion and dephasing. As the number of emitters and interactions increases, so too does the difficulty of identifying the present physical mechanisms, including sources of disorder. However, experimental access is limited to specific signatures such as time-resolved photon counts and spectra, while the underlying Hamiltonian or Liouvillian is not directly observable. Many quantum Hamiltonian learning methods determine a model once the term-by-term structure is specified, but in realistic settings the structure itself is uncertain~\citep{schmidt2009distilling, wang2017experimental, bakshi2024structure, gentile2021learning, wallace2025learning, Poteshman2025, ivashkov2026ansatz}.

Recent advances in large language models (LLMs) offer a new approach to this problem. Unlike conventional machine learning methods designed for a fixed model class or task, LLM-based agents can combine cross-disciplinary scientific knowledge from its pretraining with literature search, code generation, and computational tools to reason over a diverse space of hypotheses~\citep{lewis2020retrieval, hendrycks2020measuring, chen2021evaluating, lu2025octotools}. These capabilities are particularly relevant when the physical model structure itself is unknown, since the agent can propose alternative Hamiltonian or Lindbladian terms, implement the corresponding models, and evaluate them against experimental data without requiring a separately engineered workflow for each hypothesis. Although LLM agents have begun to demonstrate such open-ended scientific reasoning in other domains, their use for autonomous Hamiltonian or Lindbladian model discovery from quantum experiments remains largely unexplored~\citep{m2024augmenting, lu2024ai, ghafarollahi2025automating, swanson2025virtual, mitchener2025kosmos, wang2025spatialagent, alber2026cellvoyager, zhang2026virtual, nagele2026agentic}.

Here we introduce an artificial intelligence (AI) driven model discovery workflow for quantum few-body systems that autonomously extracts Hamiltonian/Lindbladian descriptions from experimental data. Our method employs a large language model (LLM) agent equipped with predefined skills and a structured workflow to propose candidate models and evaluate them. The architecture requires only a single experiment-specific document from the human scientist. The rest of the workflow, which includes literature search, candidate model selection, research plan generation, Python tool generation, and analysis execution, are conducted autonomously by the agent. At the same time, the workflow is designed for optional human oversight and involvement as a pathway for human-AI collaboration. Applied to experimental datasets collected in our laboratory from disordered emitters coupled to nanophotonic cavities, the agent generates analyses that confirm previous human expert analyses and provides further insight about the specific parameters of disorder. Finally, we assess the reproducibility of the agentic analysis by repeating the workflow from the same human-provided input and comparing the results across independent runs.

% \subsection{Related works}
% Related works.

\section{System Architecture}
Applying a general-purpose LLM agent directly to a scientific problem can make it difficult to trace and evaluate the scientific reasoning and intermediate decisions underlying its conclusions, potentially obscuring methodological errors even when results appears plausible~\citep{turpin2023language, bulian2023assessing}. This motivates the implementation of structured workflows, external scientific knowledge, and computational tools to support and validate AI agents in scientific tasks~\citep{m2024augmenting, chen2025scienceagentbench}. Our framework addresses these limitations through a structured three-stage workflow consisting of hypothesis and research plan generation, computational tool generation, and analysis execution. Organizing the analysis into explicit stages with well-defined intermediate outputs reduces task complexity and provides natural checkpoints for human oversight. The agent is implemented in Claude Code (Claude Sonnet 4.6), which provides built-in functionality for code generation and execution, web search, and custom skill use~\citep{anthropic2025claudecode, anthropic2026sonnet46}.

\begin{figure}[t]
\begin{center}
%\framebox[4.0in]{$\;$}
\includegraphics[width=0.8\textwidth]{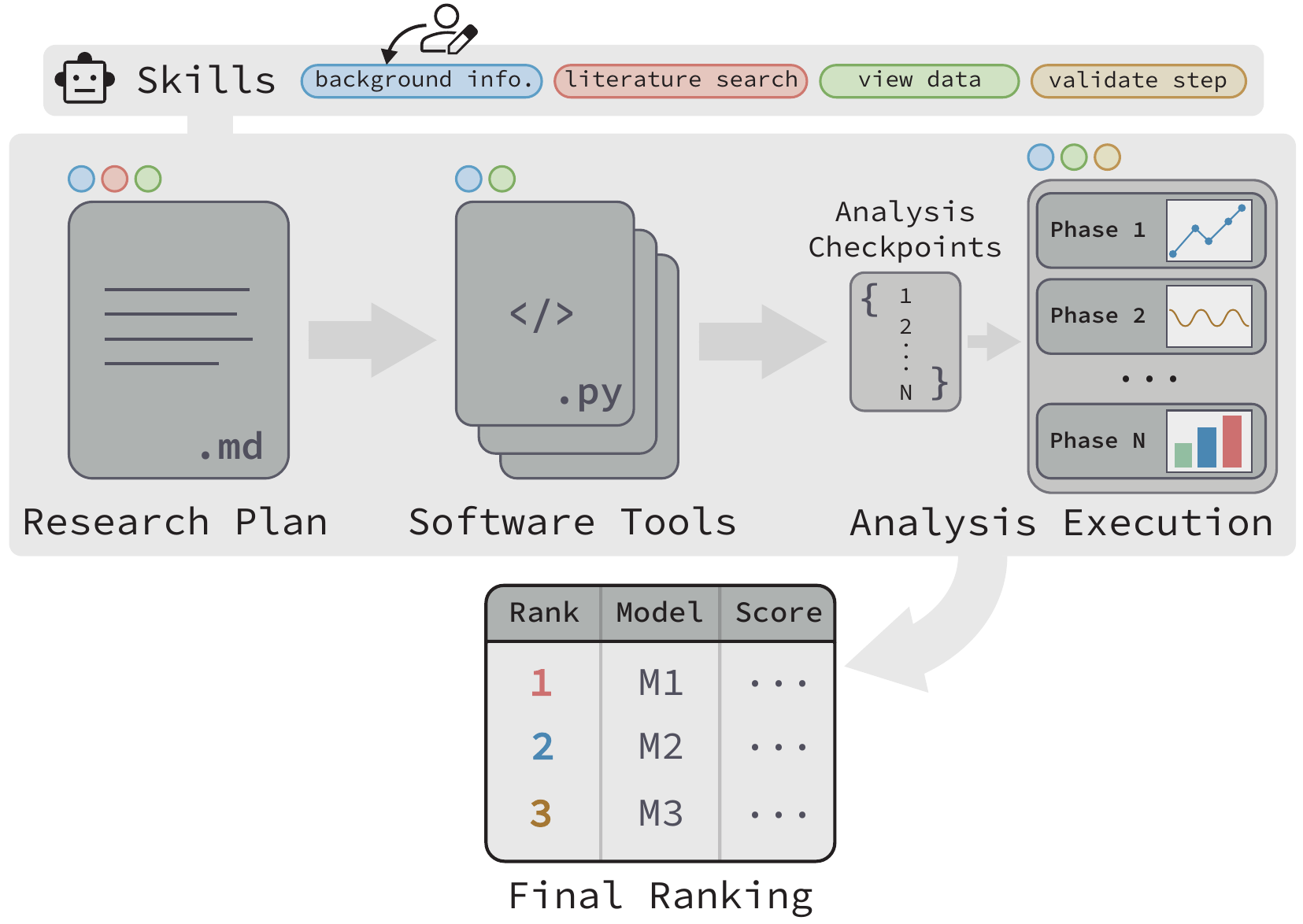}
\end{center}
\caption{\textbf{Agentic AI framework for discovering physical models from experimental data.} Given a single \texttt{experimental-background} document that outlines the experiment and broad analysis goals, and equipped with the necessary skills, the agent autonomously carries out each step: research plan generation, computational tools generation, then analysis execution. Finally, the agent produces a ranking of all the candidate models. Artifacts like Markdown documents, Python files, simulation results, and figures are recorded throughout the process. The structured workflow easily allows for human expert oversight or collaboration (e.g., using their own simulation code).}
\label{fig:workflow}
\end{figure}

\paragraph{Scientific workflow.} Beyond the workflow scaffolding, the agent requires only a single experiment-specific document describing the laboratory setup, experimental data, physical constraints, and analysis goals as its input. This document is the only component that changes when the workflow is applied to a new experiment, making the architecture readily transferable across experimental settings. From this input, the agent proceeds autonomously through the following three-stage workflow (Fig.~\ref{fig:workflow}):

\begin{enumerate}
    \item Research plan generation.
    \item Computational tool generation.
    \item Analysis execution.
\end{enumerate}

% This structure serves two purposes: it follows the logical progression of the scientific method, and it creates natural checkpoints at which the scientist can monitor progress or (optionally) intervene. 
At each stage, the agent produces machine- and human-readable artifacts, e.g. Markdown documents, Python source files, and figures, that document its reasoning and outputs. After the workflow concludes, the agent reports the findings and concludes with a ranked list of candidate models.

\paragraph{Skills.} A key component of the workflow is the agent's access to Claude Agent Skills~\citep{anthropic2025agentskills}, structured prompts that constrain and specialize its behavior but are used autonomously by the agent. The skills implemented here are:
\begin{itemize}
    \item \texttt{\textbf{experimental-background}}: stores the scientist-supplied experimental document and makes it available for consultation at any point in the workflow.
    \item \texttt{\textbf{literature-search}}: searches for and retrieves papers on a specified topic, prompting the agent to use WebSearch to identify titles then fetch full text articles via \texttt{arxiv.org/html/\{ID\}}.
    %The skill is presented as a multi-step guide in order to maximize the number of full-text articles accessed. Full-text retrieval is ideal, as equations, experimental parameters, and derivation details are often absent from abstracts.
    \item \texttt{\textbf{explore-data}}: performs a comprehensive survey of the experimental data files and reports format, dimensionality, and parameter value ranges.
    \item \texttt{\textbf{validate-phase}}: enforces a formal gate between analysis phases (Sec.~\ref{sec:workflow:analysis}) by reporting whether or not all required outputs exist, numerical results have converged, and no errors appear in the runtime log.
\end{itemize}

We describe all three stages of the workflow below. At the start of every stage, the agent invokes the \texttt{experimental-\allowbreak background} skill to ensure that all decisions are aligned with the analysis objectives.

\subsection{Plan generation}
\label{sec:workflow:plan}
The agent begins with an extensive literature search, using the \texttt{literature-search} skill to retrieve full-text papers and abstracts, which enables the agent to identify and motivate the candidate Hamiltonian/Lindbladian model space. The literature search also serves to inform the agent of various simulation and statistical model evaluation techniques that may be applicable for the given setting. After surveying the literature, the agent formulates a research plan which outlines the candidate physical models to evaluate, quantitative model evaluation techniques, and final ranking strategy. The plan then undergoes a revision step in which the agent checks cited sources, verifies that the modeling choices are physically motivated, and corrects any errors. The research plan is saved as a Markdown document that serves as the main guiding reference for all subsequent stages.

% The plan then undergoes a revision step in which the agent checks cited sources for accuracy, re-reads the experimental background to verify that every modeling choice is physically motivated, and corrects any technical inconsistencies. The combined literature search over these two steps typically spans dozens of papers, employing the HTML-first retrieval protocol to access as much scientific content as possible.

% The resulting plan serves as both a machine-readable reference for the agent to use throughout the analysis, and a human-readable scientific document, giving the scientist a natural point of oversight before the pipeline advances.

\subsection{Tool generation}
\label{sec:workflow:tools}
In the second stage, the agent generates a Python tool library for executing the research plan. The library includes functions for model definition, physics simulations, statistical fitting, and visualization, which are saved to a directory and imported by the subsequent analysis scripts. Pre-generating these tools makes the computational framework explicit and inspectable, allowing a scientist to review or modify the library before data analysis begins, if desired. It also establishes the main computational methods in advance, reducing the agent's mid-analysis implementation choices and making the analysis easier to follow and reproduce. If the human scientist instead provides an existing computational library, the agent incorporates it and generates only the additional tools needed to complete the analysis.

\subsection{Analysis execution}
\label{sec:workflow:analysis}
In the final stage, the agent performs the full analysis. First, the agent generates a checkpoint list derived from the research plan to break down the research plan into sequential phases, each with explicit output requirements that serve as entry conditions for the next. For each phase, the agent reviews preceding outputs to assess progress and resolve any incomplete or non-converged results. It then writes and saves a Python script using functions from the library generated in the tool generation step, reviews the code for errors, and performs a timing probe before executing the full computation. The resulting data, figures, and summaries are saved, and the agent uses the \texttt{validate-phase} skill to assess whether the required criteria have been met before advancing (Fig.~\ref{fig:supp:execution}). 

% The following procedure is repeated for all five analysis phases:

% \begin{enumerate}
%     \item \textbf{State audit}. The agent reviews all preceding phase outputs to assess current progress. If any outputs from the preceding phase are incomplete or non-converged, the agent resolves them before advancing.
%     \item \textbf{Script implementation}. The agent writes and saves a Python script for the current phase. The script imports functions from the library generated in the tool generation step.
%     \item \textbf{Performance audit}. The agent reviews the code, fixing any errors, and runs a timing probe for executing one representative unit of work. This ensures adherence to a reasonable time budget before committing to the full computation.
%     \item \textbf{Execution}. The agent runs the script. All resulting data is saved, together with a summary documenting the agent's reasoning and outputs. The agent also produces figures to visualize the results.
%     \item \textbf{Validation}. The agent dispatches the \texttt{validate-phase} skill, which assesses success against all required criteria and identifies any outputs that require remediation.
% \end{enumerate}

Crucially, while the script is running, the agent actively monitors progress and identifies when physical or numerical issues arise. When the agent catches errors during the execution, it interrupts the script, edits the code, and re-executes the phase, all without human intervention. By the end of the fifth final analysis phase, the candidate models are all thoroughly evaluated and ranked. In addition to this final ranking, each analysis phase records useful artifacts, such as simulation results, figures, and summaries that supplement the main findings.

Scientific analyses often conclude with unresolved questions or new hypotheses. To address this, our workflow optionally allows the agent to initiate a follow-up investigation by reviewing its results and original goals, performing additional literature search, and identifying a useful next direction. The follow-up may involve testing new hypotheses, increasing computational effort, or further distinguishing between leading candidate models. It then repeats the full three-stage workflow, generating a new research plan, computational tools, and analysis based on the preceding results.

\section{Discoveries from Laboratory Experiments}
\label{sec:results}
We apply the model discovery workflow to three previously published cavity-QED experimental datasets from our laboratory, spanning increasingly complex physical systems. Because each dataset was previously analyzed by human researchers, they provide a means for comparison between AI and human expert interpretations. Across the three experiments, the agent recovers the established physical understanding while extending the original analyses by identifying unresolved emitter parameters, discriminating among competing physical mechanisms, and quantifying multiple sources of disorder.

\subsection{Two SnV$^-$ Color Centers in a Photonic Crystal Cavity}
\label{sec:exp2}
We first applied the framework to experimental data from SnV$^-$ color centers coupled to a one-dimensional photonic crystal cavity in diamond (Fig.~\ref{fig:exp2}a), which was reported in Ref.~\citet{lee2026quantum}. SnV$^-$ centers are an emerging solid-state qubit platform for quantum networks and quantum information, and integration with nanophotonic cavities provides a route to enhanced light–matter interaction and improved qubit fidelity. 

When a cavity is resonant with one of the optical transitions of an emitter, either the C or D transition, the spontaneous emission rate is enhanced through the Purcell effect, producing a reduction in the measured excited-state lifetime. This lifetime reduction is the primary observable used to extract the Purcell factor, which serves as the figure of merit for quantifying cavity-enhanced light–matter interaction in this weakly-coupled system~\citep{faraon2011resonant, zhang2018strongly}. However, extracting this information from disordered multi-emitter data is challenging. For instance, weakly coupled or dim emitters may produce only subtle perturbations to the decay dynamics, especially when their signal is obscured by a brighter emitter coupled to the same cavity mode. In the present experiment, we measured the excited-state lifetime using pulsed time-resolved photoluminescence (PL), measuring the decay of light emitted from the system following an excitation pulse. Measurements were collected across a wide range of cavity detunings to probe emitters at different frequencies. A dominant emitter had been clearly identified by previous human analysis (although this fact was not made available to the AI agent), while additional weaker contributions from a second emitter were suspected but difficult to quantify using the conventional approach of fitting individual lifetime traces in isolation~\citep{lee2026quantum}. Here we show that the AI agent successfully identifies the presence of two emitters in the same data and quantifies their contribution to the observed lifetime.

\subsubsection{Results}

\paragraph{Model space construction.} After the literature search, the agent generated a list of five candidate physical models to evaluate, mostly focused on the number of emitters (Tab.~\ref{tab:candidate_models}). All models with independent (i.e. no collective dynamics) emitters share the same underlying framework; via adiabatic elimination of the cavity field in the bad-cavity limit, the agent correctly determined that each emitter $i$ experiences a Lorentzian Purcell enhancement of its spontaneous emission rate when the cavity is on resonance with either C or D optical transition:

$$\gamma_{\mathrm{tot},i}(\omega_\text{cav}) = \gamma_0 + \frac{g^2_{C,i}\,\kappa}{(\omega_\text{cav} - \omega_{C,i})^2 + (\kappa/2)^2} + \frac{g^2_{D,i}\,\kappa}{(\omega_\text{cav} - \omega_{D,i})^2 + (\kappa/2)^2}$$

where $\gamma_0$ is the non-cavity-enhanced emission rate, $\omega_{\rm cav} - \omega_{\alpha, i}$ is the detuning of the $i$th emitter's $\alpha$ transition relative to the cavity's resonant frequency, $g_{\alpha, i}$ is the coupling strength between the $i$th emitter's $\alpha$ transition dipole and the cavity field, and $\kappa$ is the cavity loss rate. Finally, the decay shape is modeled by:

$$I(t) = \sum_i r_i \, e^{-\gamma_{\mathrm{tot},i}(\omega_\text{cav})\, t}, \qquad \sum_i r_i = 1$$

where $r_i$ are the relative photon count amplitudes.

Additionally, the agent decided to include a two-emitter model with cavity-mediated collective effects. The agent correctly identified the explicit formulas for the two-emitter collective decay rate and spin-exchange rate,
$$
\Gamma_{12}  = \frac{g_{C,1}g_{C,2}\kappa}{{\Delta}_{12}^2 + (\kappa/2)^2} \text{ and } J_{12} = \frac{-g_{C,1}g_{C,2}{\Delta}_{12}}{{\Delta}_{12}^2 + (\kappa/2)^2} \,, \quad (\Delta_{12} = \omega_{C,1} - \omega_{C,2})
$$
respectively, which appear in the master equation~\citep{evans2018photon}. An upper bound on the number of emitters $N \leq 4$ for the candidate models was derived by the agent using the \texttt{explore-data} skill; the subagent identified four potential spectral peaks in the off-resonant PL spectrum within the cavity sweep range.

\begin{figure}[!t]
\begin{center}
%\framebox[4.0in]{$\;$}
\includegraphics[width=1\textwidth]{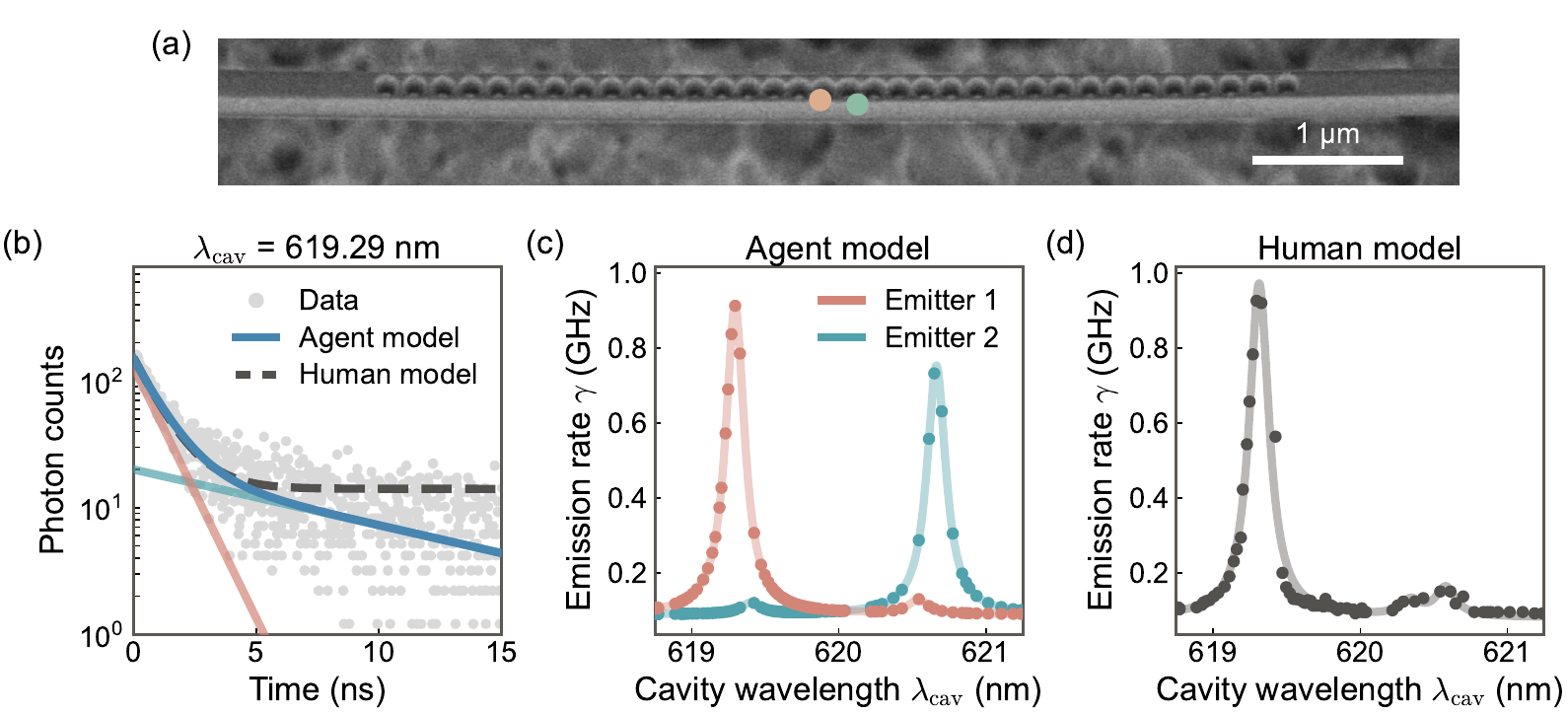}
\end{center}
\caption{\textbf{AI agent determines a model for two disordered SnV$^-$ color centers in a diamond cavity.} \textbf{(a)} Scanning electron microscope (SEM) image of a diamond photonic crystal cavity. Two representative emitters (circles) are drawn for illustrative purposes; they are embedded in the diamond and not visible in the SEM. \textbf{(b)} One measurement from the dataset of lifetime data (time-resolved photoluminescence), showing the AI agent's model improvement over the human model in representing the data. The full dataset can be found in Figure~\ref{fig:supp:exp2}. \textbf{(c)} Spontaneous emission rate as described by the AI agent's model, clearly delineating the contributions from both emitters. \textbf{(d)} Spontaneous emission rate as described by the human model, which only approximates the contribution of the second emitter. For (c) and (d), each dot represents a lifetime measurement; the value is the corresponding fitted spontaneous emission rate. In (d), the line is a quadruple Lorentzian fit to the points.}
\label{fig:exp2}
\end{figure}

\paragraph{Analysis.}
Drawing on prior work in quantum Hamiltonian learning~\citep{wang2017experimental,gentile2021learning,Poteshman2025}, the agent adopted a Bayesian model selection framework. Model parameters were optimized jointly across all experimental traces using differential evolution to identify the best-fit parameter values~\citep{storn1997differential}. Candidate models were first screened using the Bayesian information criterion (BIC), after which the leading models were ranked by their Bayesian model evidence evaluated by nested sampling~\citep{schwarz1978estimating, skilling2006nested}.

With the candidate models and model selection procedure established, the agent first performed a simple analysis of the dataset by independently fitting each of the 110 lifetime traces to a single exponential to identify trends. This revealed a lifetime reduction when the cavity was tuned to various peaks in the photoluminescence spectrum, suggesting the presence of at least one emitter and a possible second, weaker emitter. This initial approach is similar to the previous human analysis. The agent then proceeded to evaluate the candidate physical models jointly across the full dataset. During the analysis, the agent found that the three- and four-emitter models had higher Bayesian evidence than the two-emitter model. The agent realized the fitted lifetimes were shorter than the instrument response time, however, and disqualified them.

\paragraph{Findings.}
A summary of the analysis breakdown can be found in Figure~\ref{fig:supp:execution}. The agent identified two independent SnV$^-$ color centers:

\begin{itemize}
    \item \textbf{Emitter 1}: C transition wavelength $\lambda_C = 619.29~\rm nm$, coupling factor $g_C/2\pi = 2.0 \pm 0.02~\mathrm{GHz}$, D transition wavelength $\lambda_D = 620.54 \pm 0.02~\rm nm$, corresponding to a C transition Purcell factor $F_\text{Purcell} = 22.8 \pm 0.5$.
    \item \textbf{Emitter 2}: C transition wavelength $\lambda_C = 619.40 \pm 0.01~\rm nm$, coupling factor $g_C/2\pi = 0.39 \pm 0.04~\mathrm{GHz}$, D transition wavelength $\lambda_D = 620.65 \pm 0.03~\rm nm$, corresponding to a C transition Purcell factor $F_\text{Purcell} = 1.83 \pm 0.16$.
\end{itemize}

%contributing 86.7\%
%contributing 13.3\%
% The agent exhibited nine autonomous self-correction events during execution. The most consequential occurred mid-Phase 3: differential evolution converged to a three-emitter solution with $\tau_\mathrm{eff} = 67$ ps — below the 370 ps instrument resolution limit. The agent identified the missing constraint, derived the correct coupling-strength upper bound analytically, added it to the shared tool library (\texttt{model\_fitting.py}), and re-ran the full phase. Further events included diagnosing a step-size collapse in the \texttt{rslice} nested sampler at hard prior boundaries (switching to \texttt{rwalk}), a library API incompatibility in the random-seed argument, and two multiprocessing failures from platform-specific spawn behavior — all corrected autonomously before affecting results.

Overall, the agentic analysis confirms the physical picture established by previous human analysis while also providing a more complete picture of the second, more weakly coupled emitter (Fig.~\ref{fig:exp2}b--d). We attribute this advantage primarily to the agent's global fitting strategy, which exposes trends across the cavity detuning dataset that proved difficult to identify from individual measurements.

% \subsection{Implications}

% The device contains two emitters rather than one, reducing effective single-emitter cooperativity. For quantum networking applications requiring high-fidelity single-photon emission, spectral filtering or device selection for single-emitter coupling is indicated. For analogue quantum simulation via cavity-mediated spin exchange, the finding that emitters must be within $\sim\kappa$ in transition frequency to sustain collective interactions provides a concrete engineering target — one that emerges directly from the identified model and would not be available from conventional ad hoc exponential fitting.

% The complete pipeline — from plan generation through physical validation — required approximately \$15 in API costs and under two hours of wall time on a standard laptop, with the nested-sampling phase accounting for the majority of both.

\subsection{Single V$_{\text{Si}}$ Color Center in a Resonator}
\label{sec:exp3}

We now apply the same approach to analyze a previously published experimental data set for a silicon-vacancy (V$_\text{Si})$ color center embedded in a 4H silicon carbide (4H-SiC) whispering gallery mode (WGM) disk resonator. The analyzed experimental data set is from Ref.~\citet{lukin2025mesoscopic}. This system provides a promising platform for cavity QED and quantum-photonic applications, combining optically addressable spin states and long spin coherence persisting to room temperature while also supporting high-quality, scalable nanophotonic integration~\citep{castelletto2020silicon, lukin20204h}.

Our goal is to characterize the emitter--cavity interaction and extract the coupling strength $g$ of a single color center from a single lifetime measurement. In the strong-coupling regime $(g>\kappa/4)$, time-resolved photon emission following an above-resonant excitation pulse can exhibit Rabi oscillations. Experimentally, however, such oscillations were not observed; instead, the lifetime exhibited an unusual biexponential decay~\citep{lukin2025mesoscopic}. Previous human analysis attributed the lack of coherent oscillations despite $g>\kappa/4$ primarily to spectral diffusion arising from shot-to-shot fluctuations of the emitter transition frequency~\citep{lukin2025mesoscopic, de2021materials, van2025check}. However, several physical mechanisms can produce overlapping signatures within a single lifetime trace, and different combinations of these effects can describe the data similarly well. Distinguishing their relative contributions is therefore important both for understanding the observed dynamics and for identifying the dominant sources of disorder in the system.

\subsubsection{Results}

\paragraph{Model space construction.} Unlike in the SnV$^-$ experiment, the present dataset is limited to one lifetime measurement and therefore supports many plausible physical explanations. The agent constructed a space of candidate models by combinatorially including or excluding individual mechanisms (Tab.~\ref{tab:candidate_models}). It included the effects considered in the human analysis (two optical transitions, WGM basckattering, and Gaussian-sampled spectral diffusion). The agent expanded this model space by also including a Lorentzian distribution for the spectral diffusion, as well as Markovian pure dephasing and intersystem crossing (ISC) through a metastable dark state. This resulted in 32 candidate models spanning different combinations of these effects.

Because the cavity cannot be adiabatically eliminated in the strong coupling regime, the agent computed the full quantum dynamics. Each candidate model begins with a two-level emitter coupled to one optical cavity mode through the Jaynes-Cummings Hamiltonian,

$$
\hat H =
\Delta \hat\sigma^+\hat\sigma^-
+
g\left(\hat a^\dagger\hat\sigma^-+\hat a\hat\sigma^+\right),
$$

where $\Delta$ is the emitter--cavity detuning and $g$ is the coupling strength~\citep{jaynes1963comparison}. The dynamics follow the Lindblad master equation,

$$
\dot\rho =
-\frac{i}{\hslash}[\hat H,\rho]
+
\sum_k
\left(
\hat L_k\rho\hat L_k^\dagger
-\frac{1}{2}
{\hat L_k^\dagger\hat L_k,\rho}
\right),
$$

where $\rho$ is the density matrix, and the jump operators $\hat L_k$ describe dissipative processes. Every candidate model includes cavity photon loss $(\hat L_k = \sqrt{\kappa}\,\hat a)$ and spontaneous emitter decay $(\hat L_{\gamma} = \sqrt{\gamma_0}\,\hat\sigma^-)$, while additional Hamiltonian terms or jump operators are added for additional physical effects depending on the candidate. For each model, the agent numerically evolved the master equation to calculate the cavity emission and mapped it onto the measured lifetime counts.

\paragraph{Analysis.} The agent selected the corrected Akaike information criterion (AICc) to rank the candidate models~\citep{akaike1987factor, hurvich1989regression}. To optimize each model, it first used Latin hypercube sampling to explore the parameter space and then refined the best initial points with L-BFGS-B~\citep{mckay2000comparison, zhu1997algorithm}. To quantify the contribution of individual physical mechanisms beyond the ranking of complete models, we instructed the agent to devise a visualization of the candidate-model results. The agent constructed an ``effect importance matrix,'' closely related to established information-theoretic model-selection approaches based on matched AICc comparisons~\citep{galipaud2017farewell}. For each effect $e$, the agent identified every pair of candidate models that differed only in whether $e$ was included. If $m_{e}$ denotes the model that contains the effect and $m_{\neg e}$ denotes its otherwise identical counterpart without it, the agent defined the matched-pair $\Delta\mathrm{AICc}$ as

$$\Delta\mathrm{AICc}(m_e) = \mathrm{AICc}(m_{\neg e})-\mathrm{AICc}(m_e),$$

such that positive values indicate that the effect improves the likelihood by more than the penalty incurred for its additional free parameters. It then averaged this quantity over all available matched pairs,

$$\text{Avg.} \; \Delta \text{AICc}(e)=\frac{1}{|M_e|}\sum_{m\in M_e}\Delta \mathrm{AICc}(m_e),$$

to measure how consistently each physical mechanism improves the model across all the candidate models. Values near zero indicate little systematic preference for including the effect, whereas larger positive values indicate stronger support for its inclusion.

\begin{figure}[t]
\begin{center}
%\framebox[4.0in]{$\;$}
\includegraphics[width=1\textwidth]{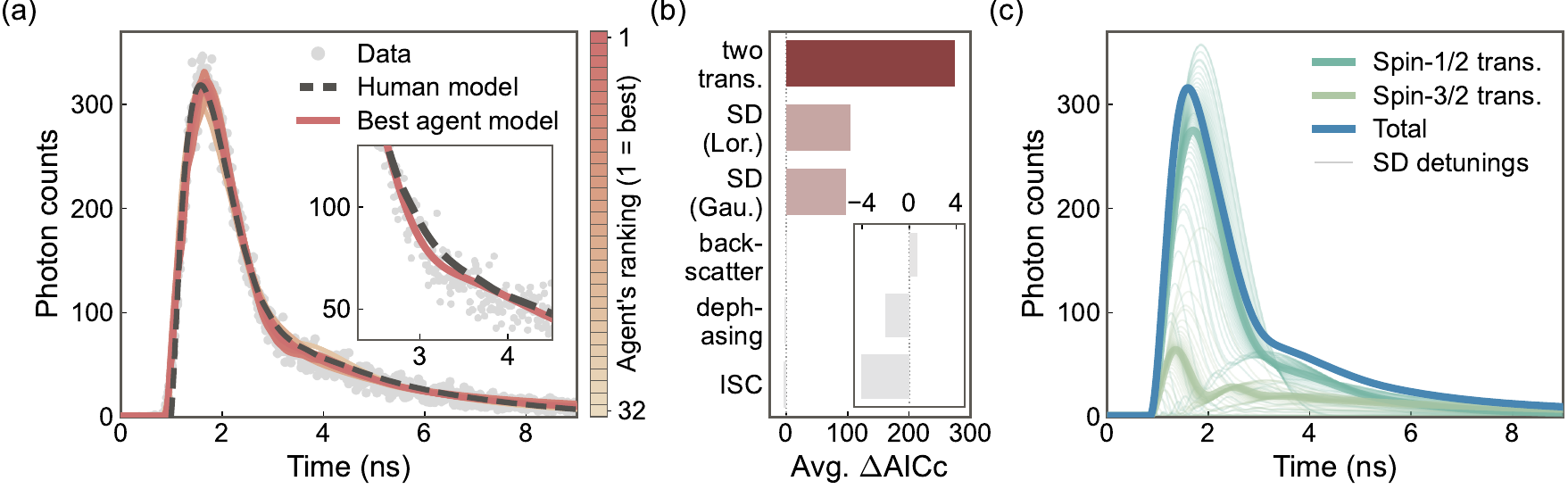}
\end{center}
\caption{\textbf{AI agent determines a model for a V$_{\text{Si}}$ color center in a silicon carbide resonator} by analyzing the experimental dataset from Ref.~\citet{lukin2025mesoscopic}. \textbf{(a)} All 32 of the agent's candidate models overlayed onto the lifetime measurement data and previous human solution. Inset shows the experimental data's bend around 3 ns, suggesting the presence of disorder. \textbf{(b)} Agent's novel effect importance analysis, showing that the two optical transitions and spectral diffusion are essential to explain the data, while mode backscattering, pure dephasing, and the intersystem crossing are not required. Note that three non-converged models are excluded from this plot. \textbf{(c)} The agent's top model (two optical transitions + Lorentzian-distributed spectral diffusion), resulting in coupling $g/2\pi = 171 \pm 1~\rm MHz > \kappa/4$. The two optical transition components and SD traces combine to form the final physical model.}
\label{fig:exp3}
\end{figure}

\paragraph{Findings.}
Figure~\ref{fig:exp3}b-d summarizes the results. Nearly all 32 candidate models fit the lifetime data and yield consistent emitter--cavity coupling strengths. The top-ranked model gives $g/2\pi=171 \pm 1~\mathrm{MHz}$, with an average value of $173 \pm 5~\mathrm{MHz}$ across the models containing both optical transitions, all above the strong coupling threshold as in previous human analysis. Because many models fit the data comparably well, the effect-importance analysis is particularly informative (Fig.~\ref{fig:exp3}c). It identifies the two optical transitions and spectral diffusion as the dominant contributions, reinforcing the central conclusions of the previous human analysis. In contrast, although WGM backscattering had been considered as a possible contributor to the biexponential decay, the agent found little support for its inclusion, due to its degeneracy with the emitter-cavity detuning parameter. Pure dephasing and the metastable dark state were similarly unsupported by this particular dataset.

In addition to the Gaussian distribution of spectral diffusion assumed in the human analysis, in the followup stage the agent also included a Lorentzian model. Gaussian fluctuations are associated with the cumulative effect of many weak perturbations, whereas the heavier tails of a Lorentzian distribution could indicate sparser, larger jumps produced by a smaller number of strongly influential charges or be a signature of long-range dipolar interactions~\citep{zumofen1994spectral, brunner2009coherent, van2025check}. The two distributions perform comparably, indicating that the present single measurement cannot distinguish between them. Nevertheless, by introducing this alternative, the agent broadened the hypotheses considered and identified a question that can motivate future measurements.

% Overall, the agent recovered the central conclusions of the human analysis while substantially expanding the physical information extracted from the same measurement. It autonomously identified additional mechanisms to evaluate, developed a unique analysis to quantify their contributions, and narrowed the likely sources of disorder in the system.

\subsection{Multiple Disordered Color Centers in a Resonator}
\label{sec:exp4}
\begin{figure}[!t]
\begin{center}
\includegraphics[width=1\textwidth]{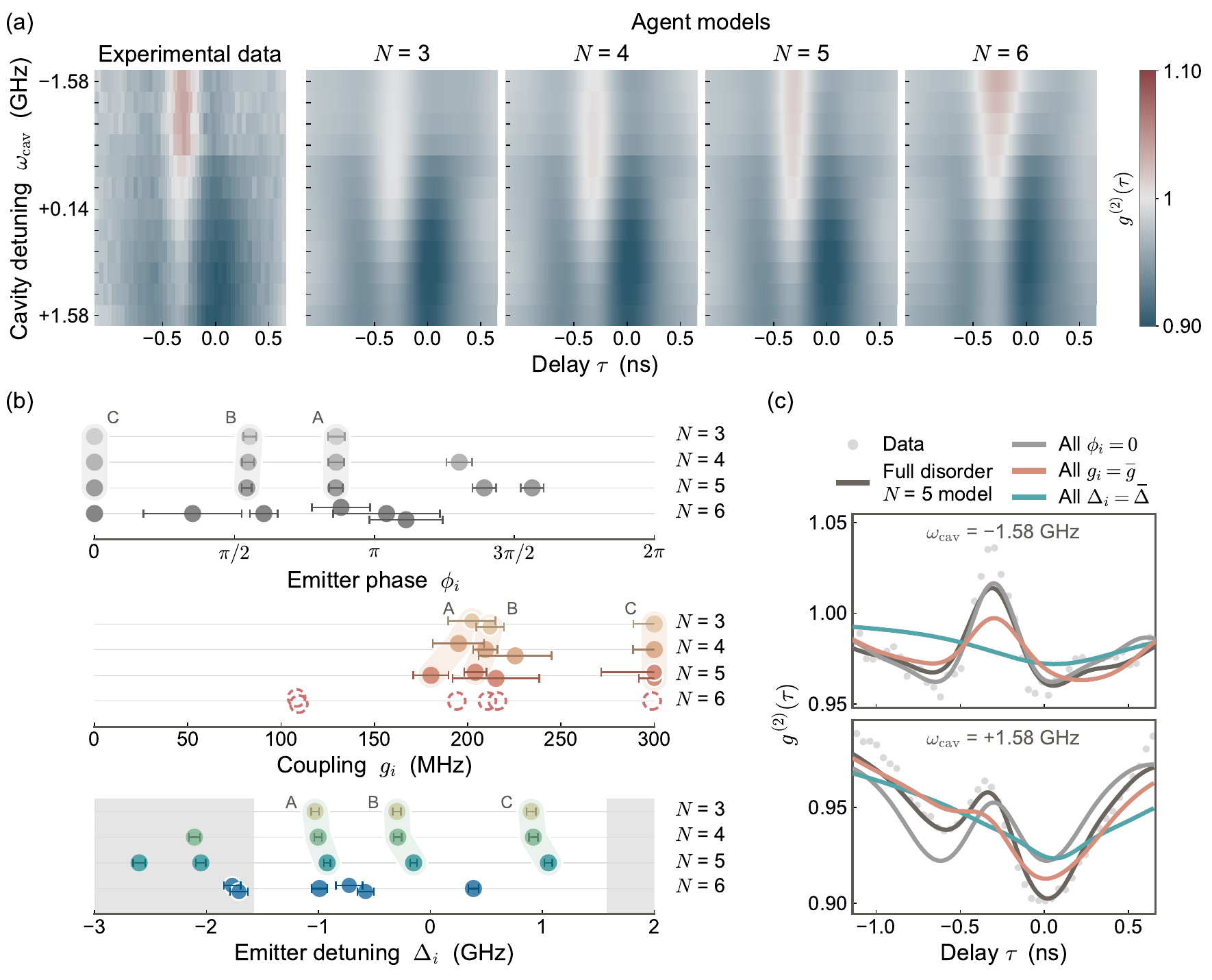}
\end{center}
\caption{\textbf{AI agent determines models for the multi-V$_\text{Si}$-cavity system} from experimental data reported in Ref.~\citet{lukin2025mesoscopic}. \textbf{(a)} Experimental cross-correlation $g^{(2)}(\tau)$ across twelve cavity detunings (left panel, adapted from~\citet{lukin2025mesoscopic}), and the agent's best-fit models with varying emitter number $N$ (four right panels). Even as small as $N=3$ emitters can reproduce the asymmetry ($\tau \sim -0.5$ to $0~\rm ns$) near the high-frequency end of the cavity scan, but the bunching near the low-frequency end of the cavity scan ($\tau \sim 0.3~\rm ns$) is only achieved with the larger-$N$ models. \textbf{(b)} Per-emitter parameters from the agent models for comparison across models with varying emitter count $N$, showing azimuthal phase disorder (top), coupling disorder (middle), and spectral disorder (bottom). The shaded regions in the emitter detuning plot denotes frequencies outside of the cavity scan range. Error bars convey statistical error only (Sec.~\ref{sec:supp:errors}); the $N=6$ solution yielded exceedingly large uncertainties in the coupling $g$ and are therefore omitted. Three emitters appear consistently across the $N=3$, $N=4$, and $N=5$ solutions, labeled A, B and C for clarity. \textbf{(c)} Results of the agent's disorder study, which involved removing one source of disorder at a time from the $N=5$ model then re-fitting, displayed for the low frequency (top) and high frequency (bottom) cavity positions. The agent found that all sources of disorder are required to reproduce the data.}
\label{fig:exp4}
\end{figure}

Multi-emitter cavity QED systems enable collective light–matter interactions and cavity-mediated emitter–emitter interactions beyond those accessible in single-emitter systems, with potential applications in quantum information, simulation, sensing, and networking~\citep{angerer2018superradiant, greve2022entanglement, lukin2023two, lei2023many}. However, these systems also present a substantially more complex disorder landscape, as variations associated with each emitter reshape the dynamics. Using our AI agent, we study previously published experimental data from a system of $\sim 10$ V$_\text{Si}$ emitters coupled to a WGM disk resonator~\citep{lukin2025mesoscopic}. In that experiment, the cavity resonance frequency was swept across a range of values, and cross-correlation $g^{(2)}$ measurements were acquired along the scan, resulting in the dataset shown in Figure~\ref{fig:exp4}a. Previous human analysis found that phase disorder played an important role in the pronounced asymmetric features present in the $g^{(2)}$ data. To isolate this effect while maintaining computational tractability, that analysis employed the bad-cavity approximation and restricted to $N=4$ uncorrelated emitters, resulting in a qualitative fit~\citep{lukin2025mesoscopic}. This experiment therefore presents a challenging analysis in which a complex parameter space must be explored to obtain a quantitative description of the multi-emitter dynamics.

\subsubsection{Results}
\paragraph{Model construction.}
The agent evaluated disordered, open two-mode Tavis--Cummings models~\citep{tavis1968exact} with $N=3\;\text{to}\;N=6$ emitters. All models include the main three sources of disorder for each emitter: spectral detuning \(\Delta_i\), coupling \(g_i\) and azimuthal phase \(\phi_i\) for each emitter, the latter arising from the different angular positions of the emitters within nodes of the resonator mode (Tab.~\ref{tab:candidate_models}).

We supplied the agent with a domain-specific simulation tool written by the authors of this manuscript, highlighting the flexibility of the workflow to include scientist-provided computational tools. Given the coupling strength, transition frequency, and azimuthal phase of each emitter, together with the cavity parameters, the simulation implements the Lindblad master equation and evaluates the steady-state cross-correlation function,

$$g^{(2)}(\tau) = \frac{\left\langle a^\dagger_{\rm cw}(0)a^\dagger_{\rm ccw}(\tau)a_{\rm ccw}(\tau)a_{\rm cw}(0)\right\rangle}{\left\langle a^\dagger_{\rm cw} a_{\rm cw}\right\rangle \left\langle a^\dagger_{\rm ccw} a_{\rm ccw} \right\rangle},$$

using the quantum regression theorem. For every candidate model, the agent treated spectral diffusion as static offsets of each emitter frequency over a Gaussian distribution and averaged over the samples.
% For this current study, the agent's solutions focus on the static disorder parameters without spectral diffusion.
% The agent ran all heavy computation on CPU nodes in Sherlock, the Stanford University shared high-performance computing cluster. 

\paragraph{Analysis.} The agent implemented the covariance matrix adaptation evolution strategy (CMA-ES)~\citep{hansen1996adapting} to find best-fit parameter values for each model, first performing a global survey with a reduced cavity Fock-space truncation to mitigate simulation time before completing the optimization using the full simulation. To ensure convergence for each model, the agent opted to execute at least ten distinct optimizations from independent starting points, concluding only when the best $1/3$ of results agree within a tolerance set by the number of free parameters. Throughout the process, the agent planned out CPU usage according to the compute availability and responded autonomously to computational obstacles. To assess the importance of individual disorder types, the agent removed one mechanism at a time (e.g. fixing all emitter detunings $\Delta_i$ to the same value) and re-optimized the remaining parameters. The full analysis took over two days of computation, during which the agent actively monitored the progress (Tab.~\ref{tab:cost}).

\paragraph{Findings.} The optimization of models including spectral diffusion converged only for $N=3$ due to compute limitations, and the fit worsened with increased spectral diffusion width (Fig.~\ref{fig:supp:exp4-SD}). Subsequently, the agent focused on models without spectral diffusion. It found that all three forms of static emitter disorder ($\Delta_i$, $g_i$, and $\phi_i$) are important for reproducing the $g^{(2)}(\tau)$ data, with the removal of any one effect causing a large degradation in the fit quality (Fig.~\ref{fig:exp4}c). Comparing BIC values, the agent saw that increasing the number of emitters improved the model, i.e. $\text{BIC}_{N=5} < \text{BIC}_{N=4} < \text{BIC}_{N=3}$. Further, the agent identified two comparably performing $N=4$ solutions with distinct parameter values, indicating the potential for multiple viable solutions due to the large parameter space. Overall, the agent concluded that the system likely contains at least five emitters, with additional emitters possibly needed to fully reproduce the $g^{(2)}(\tau) > 1$ bunching feature at $\tau \sim -0.3~\rm ns$. Indeed, the $N=6$ model achieved a bunching magnitude closer to the data (Fig.~\ref{fig:exp4}a), though the full $N=6$ optimization did not complete within the available compute, and the $N=5$ model was ultimately more favorable in BIC. The agent notably did not explicitly compare fitted parameter values from solutions across $N$, but our inspection reveals a consistent subset of three emitters within the cavity scan range (Fig.~\ref{fig:exp4}b). Given their consistency across the $N=3$, $N=4$, and $N=5$ solutions, these emitters may be responsible for the asymmetry observed at higher frequencies. This interpretation is consistent with the analysis presented in Ref.~\citet{lukin2025mesoscopic}, which demonstrated that the interplay between spectral and phase disorder was responsible for this asymmetry. Finally, the bunching feature appears to require additional emitters near the low-frequency end of the cavity scan, as seen in the $N=4$, $N=5$, and $N=6$ solutions.

\begin{figure}[!t]
\begin{center}
%\framebox[4.0in]{$\;$}
\includegraphics[width=0.92\textwidth]{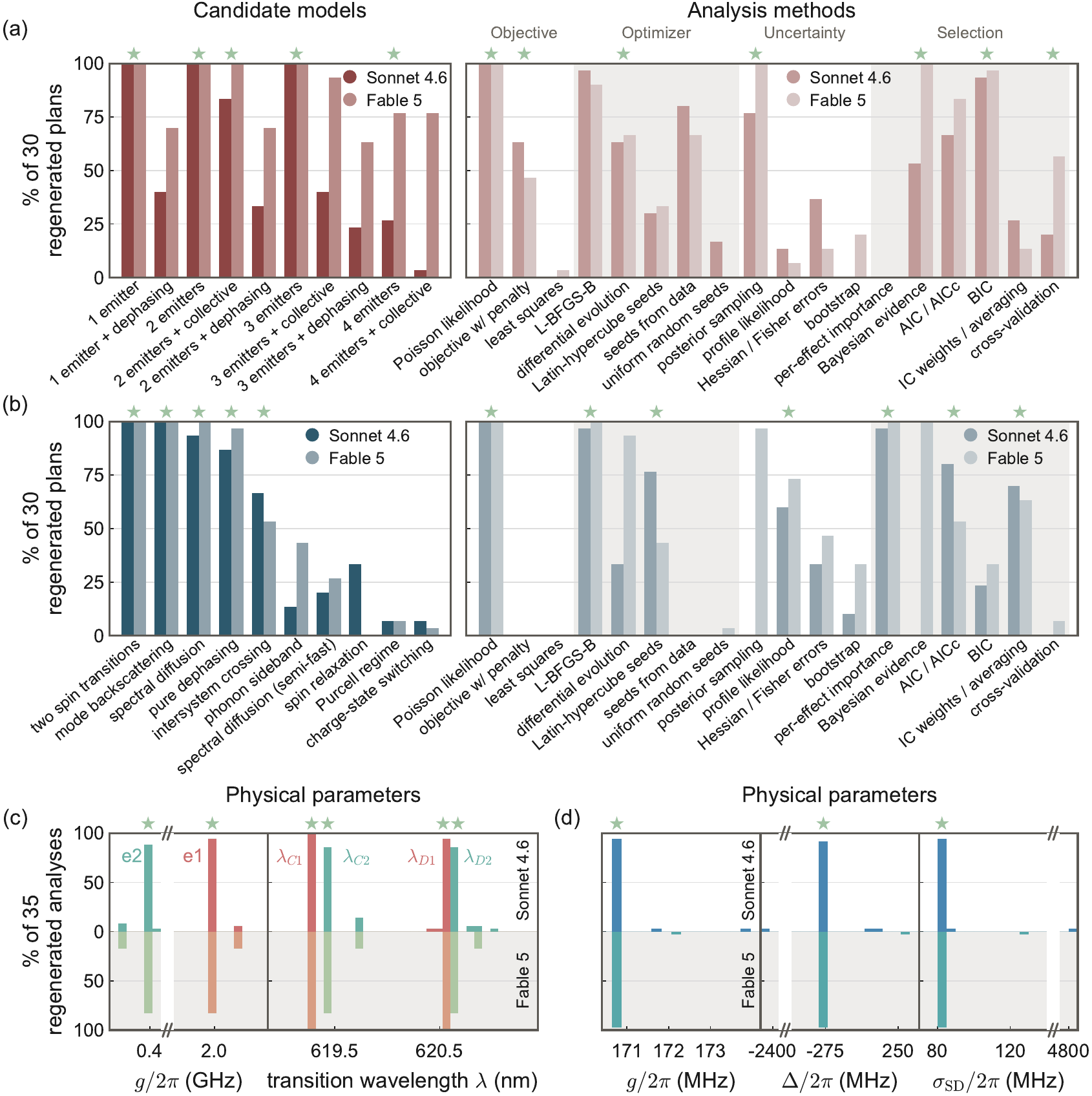}
\end{center}
\caption{\textbf{Reproducibility analysis of two different experiments}. \textbf{(a)} Percentage of specific candidate models (left) and analysis methods (right) present in 30 independently-generated SnV$^-$ research plans. \textbf{(b)} Percentage of specific candidate models (left) and analysis methods (right) present in 30 independently-generated V$_{\text{Si}}$ research plans. \textbf{(c)} Physical parameters from the winning two-emitter model in 35 SnV$^-$ analyses, showing good agreement for the coupling factor $g$ (left) and transition wavelengths $\lambda$ (right) for both emitters. Emitter 1 (emitter 2) is denoted by $e1,\; \lambda_{C1},\; \text{and}\; \lambda_{D1}$ ($e2,\; \lambda_{C2},\; \text{and}\; \lambda_{D2}$). \textbf{(d)} Physical parameters from the winning model in 35 V$_{\text{Si}}$ analyses, showing good agreement for the coupling factor $g$ (left), emitter-cavity detuning $\Delta$ (center), and spectral diffusion distribution width $\sigma_{\rm SD}$ (right) for both emitters. Note that some axes are broken for space economy and clarity. Star symbols denote the planning choices and physical results from the case studies in Sections~\ref{sec:exp2} and~\ref{sec:exp3}, most of which are shared by the majority of the regenerated plans and analyses.}
\label{fig:consistency}
\end{figure}

\section{Reproducibility Analysis}
Reproducibility is a central requirement of scientific analysis, yet LLM outputs are inherently stochastic~\citep{wang2022self, rabanser2026towards}. Here, we evaluate how consistently the agent analyzes the same input. We assess consistency at two stages of the workflow. First, given the identical human-written input document, we examine the agent's research-plan generation by measuring the similarity between candidate models and analysis strategies across independent runs. Second, holding the research plan fixed, we evaluate the consistency of the subsequent analysis by extracting the physical parameters from the winning models. To assess how the agent's behavior depends on the scientific context and information provided by the human researcher, we perform this evaluation using two distinct human-written input documents corresponding to the SnV$^-$ experiment (Sec.~\ref{sec:exp2}) and the single-emitter V$_\text{Si}$ experiment (Sec.~\ref{sec:exp3}). Finally, we repeat the evaluation using both Claude Sonnet 4.6 and Claude Fable 5~\citep{anthropic2026fable5} to assess language model variation.

\paragraph{Research plan consistency}
To assess the reproducibility of the agent's analysis planning strategies, we repeated the research plan generation step 30 independent times using the same human-provided input document. We focused on the two major components of each plan: the set of candidate models proposed for evaluation and the set of analysis strategies proposed for model fitting and comparison. For each category, we extracted the set of all elements and quantified agreement across independent plans using the mean pairwise F1 score (Sec.~\ref{sec:supp:Fscore}). Using the Claude Sonnet 4.6 agent, the mean pairwise F1 scores for candidate model enumeration were 0.64 and 0.84 for the SnV$^-$ and V$_\text{Si}$ experiments, respectively, while the corresponding scores for analysis strategies were 0.58 and 0.66. Using the Claude Fable 5 agent, the scores were 0.61 and 0.86 for candidate models and 0.66 and 0.68 for analysis strategies. Thus, independent realizations of the planning agent show substantial overlap in their scientific decisions, while also exhibiting run-to-run variation.

Examining the individual decision elements reveals a consistent core of models and analysis steps. For the SnV$^-$ experiment, one-, two-, and three-emitter models appeared in nearly all plans, while V$_\text{Si}$ plans consistently considered the two spin-conserving optical transitions, CW/CCW mode backscattering, spectral diffusion, and pure dephasing, with some variation in additional physical effects. Additionally, nearly all plans across both experiments specify at least one optimization, uncertainty quantification, and model comparison method, but vary in the particular algorithms used. A Poisson likelihood objective and L-BFGS-B local optimization were the most recurrent choices across both experiments, although the plans also adapted consistently to experiment-specific information. For example, SnV$^-$ plans frequently initialized fit parameters from the supplementary photoluminescence spectrum, whereas V$_\text{Si}$ plans typically used a data-agnostic initialization such as Latin hypercube sampling, and almost always included a method for quantifying the importance of individual physical effects. Fable-generated plans generally enumerated more candidate models, methods, and fallback strategies than Sonnet-generated plans and more frequently included full Bayesian evidence calculations, but rarely introduced effects or strategies that were not also identified by Sonnet.

Literature retrieval also varied across independent agent runs, potentially contributing to the diversity of candidate models and analysis methods considered during research-plan formulation. To quantify the literature search consistency, we recorded all the unique papers accessed in each run and computed the mean pairwise F1 agreement between the resulting sets. For Sonnet 4.6, paper agreement was 0.38 for the SnV$^-$ experiment and 0.30 for the V$_\text{Si}$ experiment; for Fable 5, the agreements were 0.41 and 0.36, respectively. Such variability is not unexpected for open-ended information retrieval, in which multiple papers may provide relevant information for the same topic. Despite the variation, we find a small consistent core in the literature retrieval, with the three most frequently accessed papers appearing in 83\% and 90\% of Sonnet-generated SnV$^-$ and V$_\text{Si}$ plans, respectively, and 87\% and 73\% of the corresponding Fable-generated plans.

\paragraph{Physical consistency}
To assess the consistency of the analysis results, we start from the same research plan and independently execute the the analysis step 35 times. We compare the key physical parameters extracted from the winning model across these independent runs to quantify analysis consistency. Here, we consider only the main analysis stage and omit the optional follow-up analysis for clarity.

For the leading two-emitter physical model in the SnV$^-$ experiment, Sonnet 4.6 and Fable 5 yielded consistent values for the relevant parameters, particularly for the brighter emitter (Fig.~\ref{fig:consistency}c ``emitter 1''). The dimmer emitter exhibited more of a bimodal spread in transition frequencies and coupling, indicating that the low signal-to-noise of this data may support more than one model for the this emitter. Nonetheless, the values were sharply concentrated, with relative inter-quartile ranges spanning less than 1.8\% of the median for all parameters using Sonnet and less than 1.5\% using Fable. In the V$_\text{Si}$ analysis, the agent showed similar consistency. While the distribution of Sonnet analyses contained outliers at large frequencies for the detuning and spectral diffusion parameters, the distributions are otherwise highly concentrated (Fig.~\ref{fig:consistency}d), with inter-quartile intervals spanning less than 0.06\% of the median for all parameters using Sonnet and below 0.02\% using Fable.

\begin{figure}[!htbp]
\begin{center}
%\framebox[4.0in]{$\;$}
\includegraphics[width=1\textwidth]{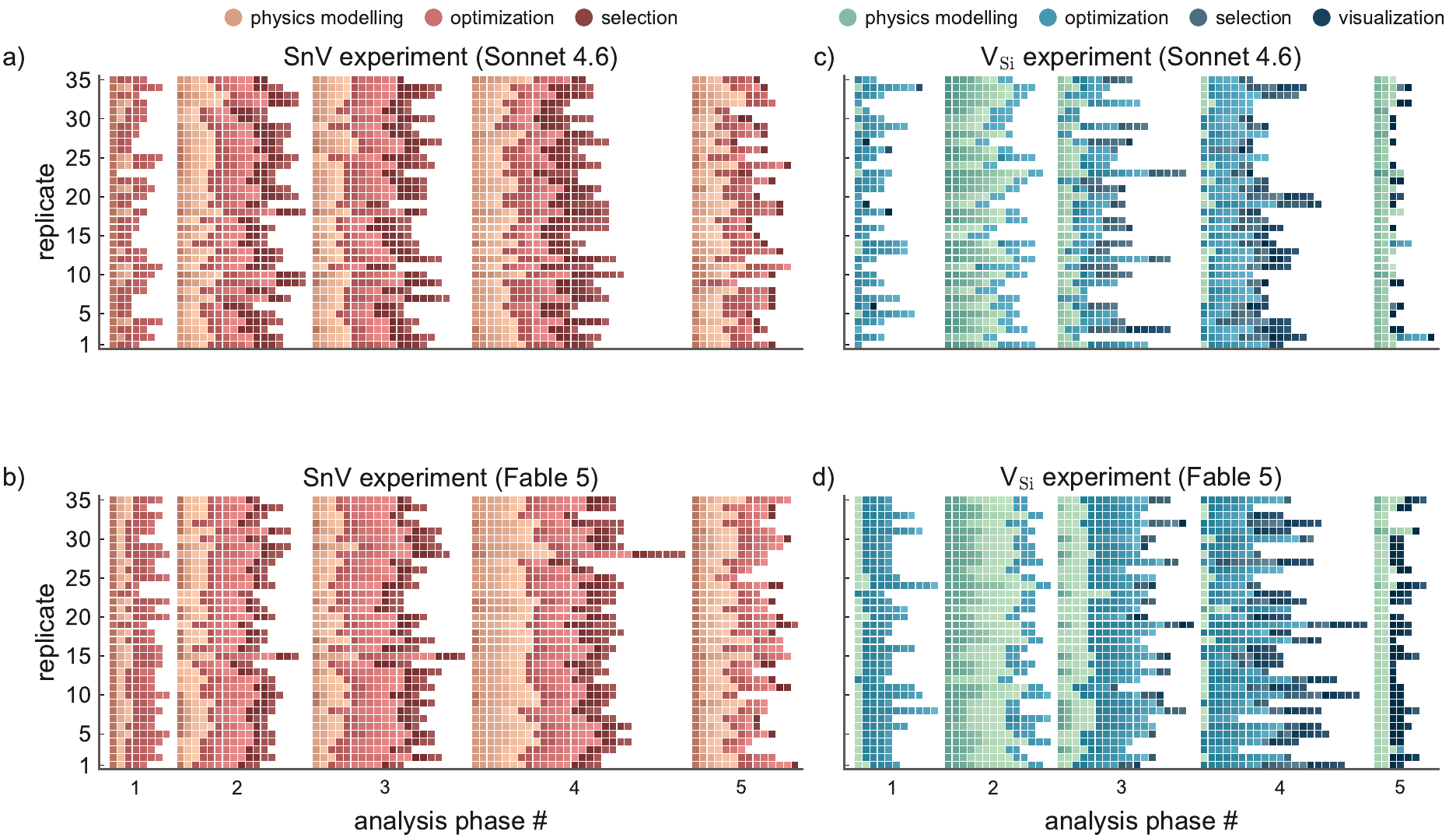}
\end{center}
\caption{\textbf{Use of the shared computational toolkit across 35 repeated analyses.} Every replicate begins from the same agent-written research plan (Sec.~\ref{sec:workflow:plan}) agent-written toolkit (Sec.~\ref{sec:workflow:tools}) and is executed with Sonnet 4.6 and Fable 5. \textbf{(a)--(b)} SnV$^{-}$ experiment, \textbf{(c)--(d)} single-V$_\text{Si}$ experiment. The horizontal axis is divided into five blocks, one per phase of the analysis (Sec.~\ref{sec:workflow:analysis}), and each square represents a toolkit function called by the agent's execution script in that phase. The functionality family is represented by color (legend), with a distinct shade per function, and follows a general phase-to-phase trend across both experiments. Claude Sonnet 4.6 and Fable 5 are closely matched on average, with 23.3 versus 27.1 distinct functions across the analysis for the SnV$^{-}$ experiment and 23.6 versus 27.6 for the V$_\text{Si}$ experiment, respectively.}
\label{fig:tooluse}
\end{figure}

These results indicate that, in contrast to the stochasticity observed during the research plan generation step, the extracted physical parameters remain relatively consistent across independent analyses given the same research plan. One feature possibly contributing to the analysis consistency is the computational library generated by the agent before execution, which provides a common set of functions through which the agent carries out the research plan. Computational tool use follows general patterns across independent analysis runs, with the agent repeatedly invoking similar functions at corresponding phases of the analysis (Fig.~\ref{fig:tooluse}).

\section{Discussion} 
Across the study of three previous experiments, the AI agent demonstrated the ability to carry out substantial insightful analysis of disordered quantum systems with limited human intervention. It identified physically meaningful parameter bounds and model equations from the literature, implemented the corresponding simulations and fitting routines, monitored their execution, and revised its approach when results became unphysical or inconsistent. In analyzing the SnV$^-$ experiment, the agent applied mid-analysis reasoning to eliminate seemingly promising but ultimately unphysical models (Sec.~\ref{sec:exp2}). In analyzing the single-emitter V$_{\text{Si}}$ experiment, the agent introduced a model-comparison strategy that quantified the contribution of individual physical effects (Sec.~\ref{sec:exp3}). In analyzing the multi-emitter V$_{\text{Si}}$ experiment, equipped with established domain software and computational resources, our agent traversed a complex and computationally demanding parameter space to yield physical insights (Sec.~\ref{sec:exp4}).

Through these studies, we identify several advantages of using structured, tool-equipped LLM-based AI agents for physical model discovery. First, the agent can identify statistical and optimization approaches that may fall outside the standard knowledge of a specialized research domain. Similarly, it can draw candidate models from the literature and test them systematically, rather than restricting the analysis to a small set of models hand-picked in advance. Finally, because the agent can autonomously implement, execute, and iterate over computational analyses, it makes model and parameter searches practical with comparatively little additional human effort (see cost and time expenditure in Table~\ref{tab:cost}).

One important limitation to using LLM-based AI agents for scientific tasks is the stochasticity inherent to LLM outputs. In our architecture, we find that this variability is concentrated in the research planning stage, where the high level approach is generally consistent but the specific set of candidate models and optimization details tend to vary between independent runs. Our comparison between Sonnet 4.6 and Fable 5 showed little difference in reproducibility. Fable 5 generally proposed a broader set of candidate models and analysis strategies, but rarely introduced physics or analysis methods that were absent from the Sonnet 4.6 runs, suggesting that more expensive models may not always be required for every step when using a structured workflow accompanied with tool use~\citep{zhou2023language, cohen2025small, li2026skillsbench}. By contrast, when starting from the same research plan, the fitted physical parameters show good agreement across runs. Consequently, one may consider refining the planning stage to improve reproducibility or to optimize for autonomous hypothesis generation in future work.

\begin{figure}[!htbp]
\begin{center}
%\framebox[4.0in]{$\;$}
\includegraphics[width=1\textwidth]{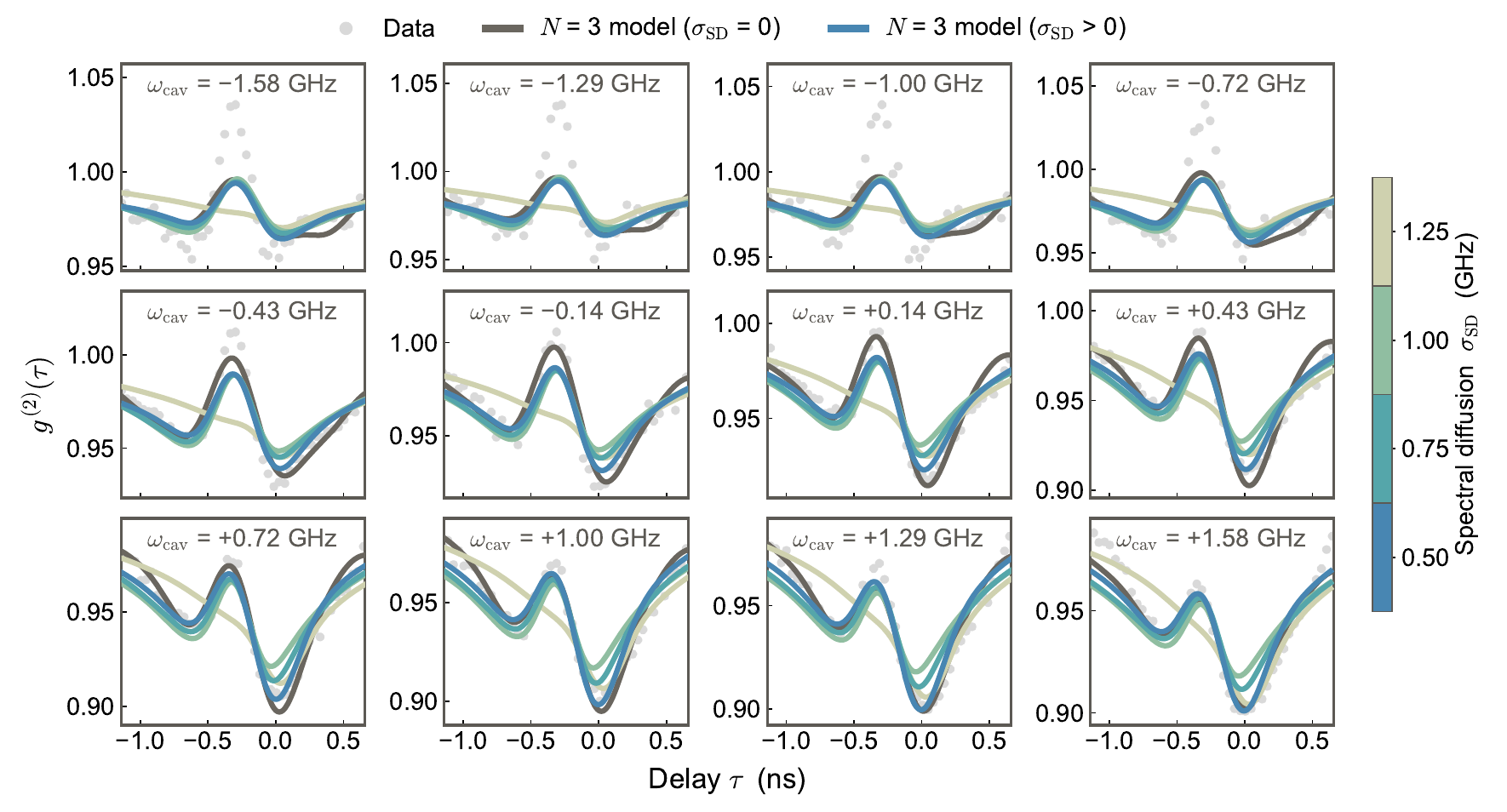}
\end{center}
\caption{\textbf{Results of the AI agent's spectral diffusion study for the multi-emitter dataset} (Sec.~\ref{sec:exp4}), shown at all 12 cavity detunings $\omega_\mathrm{cav}$. Knowing from previous experiments that spectral diffusion likely affects this system, the human-provided input document prompted the agent to consider this effect. The agent modeled spectral diffusion by shifting each emitter's transition frequency by a random Gaussian offset of standard deviation $\sigma_\mathrm{SD}$, and averaging the normalized $g^{(2)}(\tau)$ over 200 samples. Every fit was worse than the $\sigma_\mathrm{SD}=0$ case, so the agent concluded that spectral diffusion is definitely not required by the data. However, although it may not be required within this narrow definition, could it still be present? More physically-motivated tests, such as evaluating the interplay between spectral diffusion and other static forms of disorder, are needed before drawing such a concrete conclusion. This illustrates an important limitation of LLM-based agents in scientific work, namely that they are prone to stating conclusions with more certainty than the evidence warrants; as such, outputs must be reviewed against expert judgment.}
\label{fig:supp:exp4-SD}
\end{figure}

Like other LLM-based architectures, our AI agent is still susceptible to errors and limitations in scientific reasoning. For example, in the multi-emitter experimental dataset (Sec.~\ref{sec:exp4}), the agent provided useful physical insights but also asserted that the data did not support the presence of spectral diffusion, a likely source of disorder in this system. Yet this conclusion was drawn from limited evidence (Fig.~\ref{fig:supp:exp4-SD}), owing both to computational constraints and the agent's failure to devise a more physically-informed and comprehensive evaluation. Our structured workflow helps mitigate such issues by allowing for the agent's reasoning and conclusions to be manually audited throughout the analysis. Nevertheless, human expert supervision is still essential for verifying agent outputs.

% In summary, we find that our agentic AI workflow is successful at extracting physical models and revealing insight from experimental datasets of complex disordered quantum experiments. At the same time, the presence of run-to-run variability underscores the importance of expert human oversight. Our results motivate structured and interpretable agentic workflows in which the analysis proceeds through well-defined stages with natural checkpoints for human inspection, validation, and intervention when needed. 

%\subsubsection*{Author Contributions}
%(TBD)

%\subsubsection*{Acknowledgments}
%(TBD) 

% \section{Data availability}
% All data, including experimental datasets, human input documents, and agent outputs can be found \href{https://drive.google.com/drive/folders/1h0etMM6Y3_e70ZQyb_aB55jVTRvrsavs?usp=share_link}{here}.

\section{Acknowledgments}
This work was supported by the Department of Energy, Grant No. DE-SC0025295. H. C. K. acknowledges support from the Burt and Deedee McMurtry Stanford Graduate Fellowship. The experimental data for color centers in silicon carbide were collected as part of the work supported by the Vannevar Bush Faculty Fellowship project. M. B. acknowledges support from the Spanish Research Agency (MICIU/AEI), Grant No. CEX2023-001316, and the César Nombela Talent Atraction Program of the Comunidad de Madrid, Grant No. 2025-T1/TEC-36377.
% D. C. was supported by the Vannevar Bush Faculty Fellowship from the US DoW.
We thank Dominic Catanzaro for fruitful discussions on silicon carbide experiments and analysis, Jakob Grzesik for helpful discussions on computation of second-order correlations, and Hope Lee for data collection and analysis presented in~\citep{lee2026quantum}. Some of the computing for this project was performed on the Sherlock cluster; we would like to thank Stanford University and Stanford Research Computing for providing computational resources and support that contributed to these research results. 

\bibliography{main}

\section{Appendix}

\subsection{Human-provided input documents}

\paragraph{SnV$^-$ experiment.}
We gave the agent two questions: how many emitters are present, and what is the nature of their coupling to the cavity. We provided a high-level explanation of the SnV- color center and the measurement at hand. We also include a PL spectrum, which can be used to help identify potential emitters. Summary of other information included:

\emph{Physical constraints:}
\begin{itemize}
    \item The PL spectrum caps the number of transitions at about 4.
    \item The detection band contains only C lines, so any feature outside it must be the D line of an emitter already seen in the band.
    \item Every emitter must have both a C and a D transition, with separation 800--1000\,GHz.
    \item For the cavity's angular orientation, we provided the exact angular coupling factors for C and D and the expression $g^2 = g_0\, g_{\mathrm{pos}}^2\, g_{\mathrm{ang}}^2$, with $g_{\mathrm{pos}}$ equal for C and D of the same emitter.
\end{itemize}

\emph{Guidance on fitting:}
\begin{itemize}
    \item Handle the ${\sim}370$\,ps instrument response by truncating the first ${\sim}500$\,ps.
    \item Normalize the data.
    \item Fix the free-space decay rate $\gamma_0$ from the red-most traces and allow at most 20\% variation per emitter.
    \item Avoid fitting emitters to the same wavelength.
\end{itemize}

\paragraph{Single-V$_\text{Si}$ experiment.}
We asked which physical models best describe the data, whether strong coupling is present, what the coupling $g$ is, and exactly which effects are needed to explain the data. We provided a high-level explanation of the cubic V2 (k-V$_\text{Si}$) center in 4H-SiC and gave information on the microdisk and experiment at hand. We also requested the agent to provide a metric and visualization which compares the physical effects (the method is not at all specified by us). Summary of other information included:

\emph{Physical effects to consider:}
Spectral diffusion, CW/CCW backscattering, and the two spin transitions. Requested the agent to propose at least one additional effect.

\emph{Physical constraints:}
\begin{itemize}
    \item Quantum efficiencies of 0.34 (O1) and 0.64 (O2), with decay rates fixed to literature bulk values.
    \item Equal $g$ for both transitions, and spectral diffusion shifting them together.
\end{itemize}

\emph{Guidance on fitting:}
\begin{itemize}
  \item Treat excitation as instantaneous at $t_0$ and fit this quantity.
  \item Normalize the data and fit a constant count offset.
\end{itemize}

\paragraph{Multi-V$_\text{Si}$ experiment.}
We asked the agent to identify the models, the disorder, and the main causes of unusual features in a steady-state CW--CCW cross-correlation $g^{(2)}(\tau)$. We provided a high-level explanation of the V$_\text{Si}$, the microdisk, and the experiment at hand. We specified the rough $\tau$ range for the agent to analyze, in which we observe the unusual peaks and troughs in the data. Summary of other information included:

\emph{Possible sources of disorder:}
Per-emitter $g$, detuning, azimuthal phase $\phi$, and spectral diffusion.

\emph{Code provided:}
Pointed agent to our steady-state $g^{(2)}$ simulation code and helper functions for the Sherlock computing cluster.

\emph{Physical constraints:}
\begin{itemize}
  \item Emitter linewidth 65\,MHz; dephasing 40\,MHz.
  \item Bounds: $g \le 300$\,MHz, $\phi \in [0, 2\pi)$.
\end{itemize}

\emph{Guidance on fitting:}
\begin{itemize}
  \item Prioritize simulation convergence over evaluating more models.
  \item Start with a phenomenological fit over $\pm 6\,\mu$s, then fit the $\pm 2$\,ns region with arbitrary functions to find the unsual features. Consider using steps \#1 and 12 for checking models quickly to save on the full 12-step optimization time.
  \item Normalize the data and fit a $\tau_0$ offset.
  \item Pay attention to the bunching peak in measurement \#12.
\end{itemize}

\subsection{Statistical uncertainty calculation}
\label{sec:supp:errors}
\paragraph{SnV$^-$ experiment.}
Uncertainties were estimated by sampling the Bayesian posterior distribution with nested sampling, using a Poisson likelihood for the photon counts in every time bin of all 110 measurements. Each reported parameter value is the mean of the samples, and uncertainty is the standard deviation.

\paragraph{Single- and Multi-V$_\text{Si}$ experiments.} Uncertainties for both V$_\text{Si}$ experiments were estimated from the covariance matrix. Each fit minimizes a cost function that measures the misfit between the measured photon counts and the model, with each data point weighted by the inverse of its Poisson variance. For the single-V$_\text{Si}$ experiment the cost function is the Poisson likelihood of the counts in each time bin of the lifetime trace, and for the multi-V$_\text{Si}$ experiment it is the sum of the squared differences between data and model. The covariance matrix is obtained from the inverse of the cost function curvature and evaluated numerically. The standard deviation, calculated from the covariance matrix, are the uncertainties reported in Section~\ref{sec:exp3} for the single-V$_\text{Si}$ experiment and plotted as error bars in Figure~\ref{fig:exp4}b for the multi-V$_\text{Si}$ experiment.

These methods account for the statistical error of each particular solution but do not describe the larger general uncertainty of determining the canonical emitter parameters in physical reality, for which there may be more than one comparable solution. Uncertainty due to experimental instrument resolution is omitted.

\subsection{Averaged pairwise F1 score}
\label{sec:supp:Fscore}
Given the same human-provided input document, the AI agent generated 30 independent research plans specifying the candidate models and analysis methods for the subsequent analysis. An LLM (Claude Opus 5) then parsed each research plan and extracted the candidate models and analysis methods present using a fixed global vocabulary defined with guidance from a human researcher. Agreement was quantified separately for the candidate-model sets, analysis-method sets, and sets of papers retrieved during literature search, which were recorded during plan generation. For two sets $S_i$ and $S_j$, pairwise agreement was quantified using the set-level F1 score,

$$F_{i,j}=\frac{2|S_i\cap S_j|}{|S_i|+|S_j|},$$

which is equivalent to the harmonic mean of precision and recall when either run is treated as the reference~\citep{hripcsak2005agreement}. For each input document and AI model driving the agent, the final consistency score $F$ is computed as the mean of $F_{i,j}$ over all $\binom{n}{2}$ unordered pairs of the $n$ independent runs,

$$F=\frac{2}{n(n-1)}\sum_{i<j}F_{i,j}.$$

The resulting $F$ ranges from 0 to 1, where $F=1$ indicates identical extracted sets across runs and $F=0$ indicates no overlap, with intermediate values reflecting partial agreement in the selected models, methods, or retrieved papers.

%%%%%%%%%%%%%%%%%%%%%%%%%%%%%%%%%%%%%%%%%%%%%
%%%%%%%%%%%%%%%%%% FIGURES %%%%%%%%%%%%%%%%%%
%%%%%%%%%%%%%%%%%%%%%%%%%%%%%%%%%%%%%%%%%%%%%

\begin{figure}[!htbp]
\begin{center}
%\framebox[4.0in]{$\;$}
\includegraphics[width=1\textwidth]{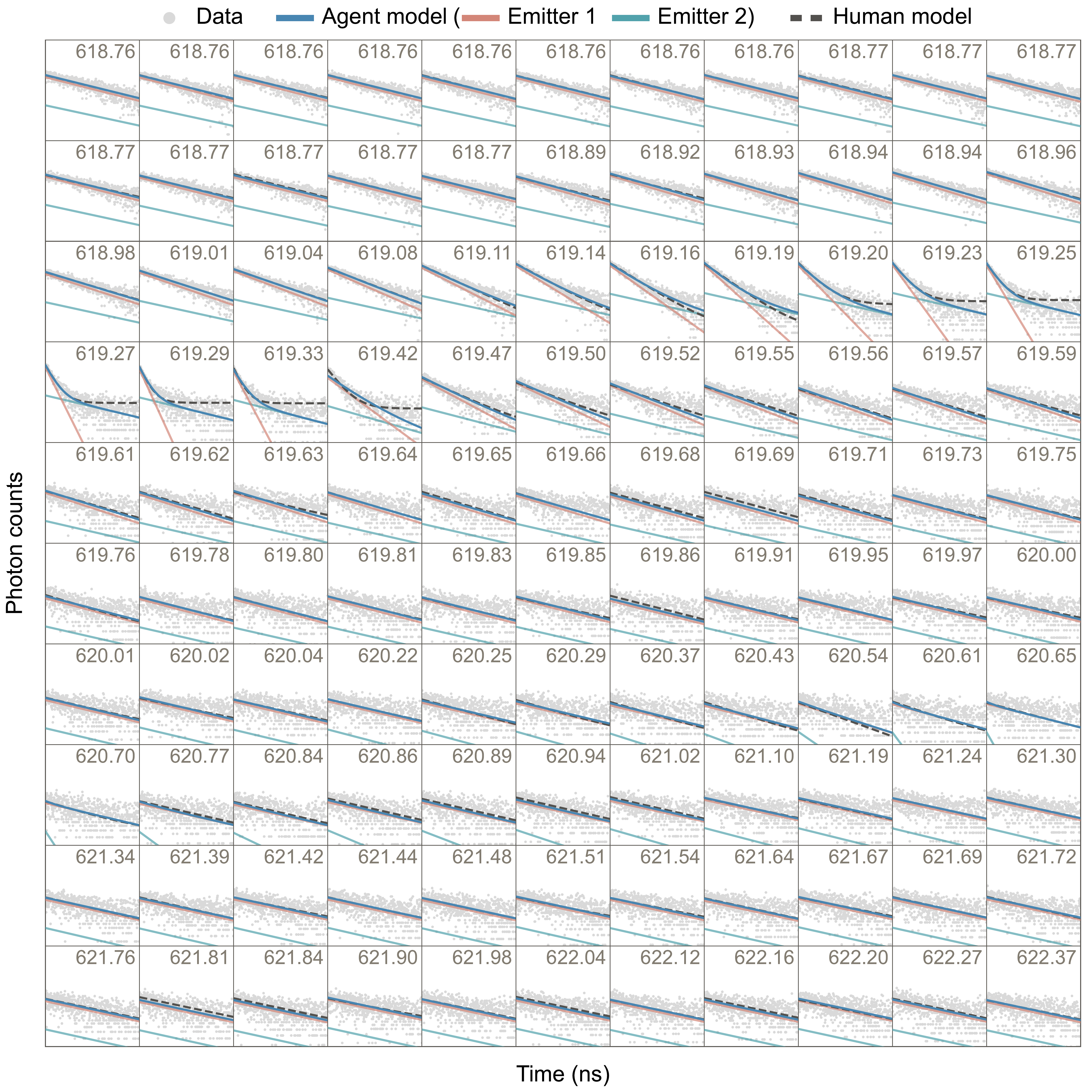}
\end{center}
\caption{\textbf{Full 110-trace lifetime dataset for the SnV experiment.} The human model (dashed) and agent's best fit two-emitter mode (blue, with per-emitter contributions in red and green) are overlaid onto the data. The x- and y-tick labels in each panel are omitted for space economy but span 0 -- 15 ns and 1 -- 800 photon counts (logarithmic scale), respectively. Each panel displays one measurement, as the cavity wavelength (upper left-hand label) is tuned by argon gas condensation from 622.37 nm to 618.79 nm.}
\label{fig:supp:exp2}
\end{figure}

\begin{table}[!htbp]
\centering
\caption{\textbf{Agent resource usage by workflow stage.} (Section~\ref{sec:results}).
\textbf{(left)} Cost in USD as reported by the Claude Agent SDK for Sonnet 4.6 usage. \textbf{(right)} Time duration, calculated with the Python \texttt{time} module. The cost of the Multi-V$_\text{Si}$ experiment is largely due to the long simulation times, during which the agent actively monitors the progress.}
\label{tab:cost}
\begin{minipage}[t]{0.48\textwidth}
\centering
\textbf{Cost (USD)}\\[2pt]
\begin{tabular}{lrrrr}
\toprule
 & Plan & Tools & Analysis & Total \\
\midrule
SnV$^-$ & 5.72 & 4.38 & 23.25 & 30.99 \\
Single V$_\text{Si}$ & 11.13 & 3.82 & 32.83 & 47.78 \\
Multi-V$_\text{Si}$ & 9.82 & 6.10 & 209.49 & 225.41 \\
\bottomrule
\end{tabular}
\end{minipage}
\hfill
\begin{minipage}[t]{0.48\textwidth}
\centering
\textbf{Time (hr)}\\[2pt]
\begin{tabular}{lrrrr}
\toprule
 & Plan & Tools & Analysis & Total \\
\midrule
SnV$^-$ & 0.74 & 0.65 & 3.76 & 5.15 \\
Single V$_\text{Si}$ & 0.98 & 0.53 & 7.27 & 8.78 \\
Multi-V$_\text{Si}$ & 0.9 & 0.6 & 58.1 & 59.6 \\
\bottomrule
\end{tabular}
\end{minipage}
\end{table}

\newcolumntype{Y}{>{\raggedright\arraybackslash}X}

\begin{table}[!htbp]
\centering
\small
\renewcommand{\arraystretch}{1.2}
\caption{
\textbf{Candidate models evaluated by the agent in the three experiments.} \textbf{(top)}
For the SnV experiment, candidate models vary by number of emitters and presence of collective effects. \textbf{(middle)} For the single-V$_\text{Si}$ experiment, candidate models are
constructed by toggling the physical effects listed below, resulting in 32 models plus a null model (uncoupled emitter). The base model (coupled emitter with one optical transition and no other effects) has 5 parameters: the emitter–cavity coupling $g$, the detuning of the spin-$\tfrac{1}{2}$ transition from the cavity $\Delta_1$, the excitation onset time $t_0$, and the count amplitude $A$ and background $B$. Subsequently, the models range from 5 to 10 parameters. \textbf{(bottom)} For the Multi-V$_\text{Si}$ experiment, the first emitter phase is fixed to 0; since backscattering is omitted, this is valid under gauge invariance. All models include a free onset delay time $\tau_0$.
}
\label{tab:candidate_models}

\begin{tabularx}{\textwidth}{
    @{}
    l
    @{\hspace{4em}}
    Y
    @{\hspace{1.2em}}
    c
    @{}
}
\toprule

\textbf{Model / Effect} &
\multicolumn{1}{c}{\textbf{Description}} &
\textbf{\# Parameters} \\

\midrule

\multicolumn{3}{@{}l}{\textbf{SnV experiment}} \\
\addlinespace[2pt]

M1 &
Single emitter, with a C transition (frequency $\omega_\text{C}$, coupling $g_\text{C}$), D transition (detuning $\Delta_\text{CD}$), and off-resonant spontaneous emission ($\gamma_0$). & 4 \\

M2 &
Two independent emitters, with count amplitude ratios $r_1$ \& $~r_2=1-r_2$ and shared $\gamma_0$. & 8 \\

M2c &
Two collective emitters (collective decay $\Gamma_{12}$ and coherent exchange $J_{12}$). &
10 \\

M3 &
Three independent emitters. &
12 \\

M4 &
Four independent emitters. &
16 \\

\midrule

\multicolumn{3}{@{}l}{\textbf{Single-V$_\text{Si}$ experiment}} \\
\addlinespace[2pt]

T &
Two optical transitions rather than one;
effect is either on or off. &
$+1$ \\

SD &
Spectral diffusion in the form of static Gaussian
($\sigma_\mathrm{SD}$), static Lorentzian ($\Gamma_\mathrm{L}$), fast pure dephasing ($\gamma_\phi$), or absent. &
$+1$ \\

BS &
WGM backscattering $\beta$ of the counter-propagating cavity modes;
effect is either on or off. &
$+1$ \\

ISC &
Intersystem crossing to a shelving state with return rates
$\gamma_\mathrm{ISC}$ and $\gamma_\mathrm{ret}$; effect is either on or off. &
$+2$ \\

\midrule

\multicolumn{3}{@{}l}{\textbf{Multi-V$_\text{Si}$ experiment}} \\
\addlinespace[2pt]

M0-3 &
Three emitters, each with coupling $g$, detuning $\omega$, and phase $\phi$. &
$9$ \\

M0-4 &
Four emitters. &
$12$ \\

M0-5 &
Five emitters. &
$15$ \\

M0-6 &
Six emitters. &
$18$ \\

MSD-3 &
Three emitters, plus spectral diffusion $\sigma_{\mathrm{SD}}$. &
$10$ \\

MSD-4 &
Four emitters, plus spectral diffusion $\sigma_{\mathrm{SD}}$. &
$13$ \\

MSD-5 &
Five emitters, plus spectral diffusion $\sigma_{\mathrm{SD}}$. &
$16$ \\

M0-FC-5 &
Five emitters, plus free cavity detunings at the ten intermediate tuning steps. &
$25$ \\

\bottomrule
\end{tabularx}

\end{table}

\clearpage
% checkpoints_results/table.tex -- written by paper_plots/make_checkpoints_table.py; edit the text there.
% The SnV analysis's execution checkpoints (experiment2/meta/PLAN11) beside one
% figure the execution agent made in each phase. Read it with
% \input{checkpoints_results/table}; it needs only graphicx in the preamble.
\begin{figure*}[t]
\centering
\footnotesize
\setlength{\tabcolsep}{4pt}
\renewcommand{\arraystretch}{1.1}
% a column's width, in per mille of the text width less the column gaps
\def\ckptw#1{\dimexpr(\textwidth-8\tabcolsep)*#1/1000\relax}
% a figure, its top level with the top of the row's first text line
\def\ckptfig#1{\raisebox{\dimexpr\ht\strutbox-\height\relax}{\includegraphics[width=\linewidth]{#1}}}
\begin{tabular}{@{}p{\ckptw{25}}p{\ckptw{140}}p{\ckptw{235}}p{\ckptw{220}}p{\ckptw{380}}@{}}
\noalign{\vskip 0pt\hrule height 0.8pt\vskip 3pt}
\multicolumn{2}{@{}l}{\textbf{Phase}} & \textbf{Task} & \textbf{Required output} & \textbf{Representative figure} \\
\noalign{\vskip 2pt\hrule height 0.4pt\vskip 5pt}
\textbf{1} &
  \raggedright Data ingestion \& system characterization &
  \raggedright Load all data; fix $\gamma_0$ from far-detuned traces; survey $\tau_\mathrm{eff}(\lambda_\mathrm{cav})$ and the PL spectrum. &
  \raggedright $\gamma_0$ with $\pm20\%$ prior bounds; angular factors $\eta_{C,D}$; positions of $\tau_\mathrm{eff}$ dips and PL peaks. &
  \ckptfig{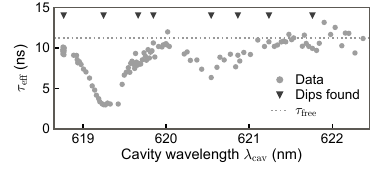} \\
\noalign{\vskip 5pt\hrule height 0.2pt\vskip 5pt}
\textbf{2} &
  \raggedright Baseline single-emitter model (M1) &
  \raggedright Build the Purcell-enhanced decay model and Poisson likelihood; fit M1 by differential evolution, then nested sampling. &
  \raggedright MAP (maximum a posteriori) fit, $\log Z$ (Bayesian evidence), BIC (Bayesian Information Criteria) and validity checks. &
  \ckptfig{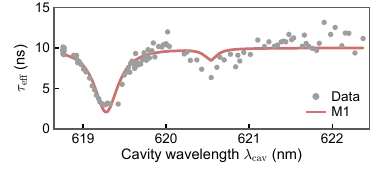} \\
\noalign{\vskip 5pt\hrule height 0.2pt\vskip 5pt}
\textbf{3} &
  \raggedright Independent multi-emitter models (M2--M4) &
  \raggedright Extend to $N=2$--$4$ emitters over all $2^N$ dipole orientations; nested-sample the best two per model. &
  \raggedright $\log Z$ and BIC for M1--M4; validity checks; fitted $\omega_{C,i}$ compared with PL peaks. &
  \ckptfig{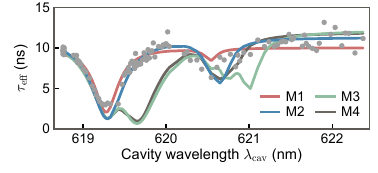} \\
\noalign{\vskip 5pt\hrule height 0.2pt\vskip 5pt}
\textbf{4} &
  \raggedright Collective decay (M2c) \& model selection &
  \raggedright Add cavity-mediated collective decay $\Gamma_{12}$ to M2; rank all models; vary $\kappa$ and the $\gamma_0$ prior to determine sensitivity &
  \raggedright Bayes factors; evidence ranking filtered by validity checks. &
  \ckptfig{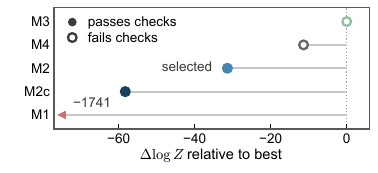} \\
\noalign{\vskip 5pt\hrule height 0.2pt\vskip 5pt}
\textbf{5} &
  \raggedright Predictive validation \& physical interpretation &
  \raggedright Test the winner on held-out traces, PL-peak alignment and C/D consistency; compare $\tau_0$ with literature. &
  \raggedright Purcell factors and $g_i$ with uncertainties. &
  \ckptfig{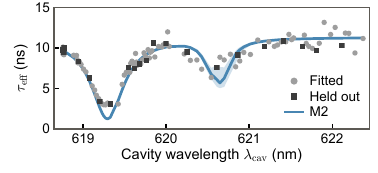} \\
\noalign{\vskip 5pt\hrule height 0.8pt\vskip 0pt}
\end{tabular}
\caption{\textbf{Summary of the SnV analysis execution.} The agent parses its research plan into five sequential phases, each with defined tasks and outputs required before the next phase (Sec.~\ref{sec:workflow:analysis}). A representative figure from each phase of the SnV analysis (Sec.~\ref{sec:exp2}) is displayed here with reformatted style and color for clarity. Due to the large quantity of data, the agent plotted the candidate models against the effective lifetime $\tau_\text{eff}$, calculated from a single-exponential fit of each decay trace.}
\label{fig:supp:execution}
\end{figure*}

\clearpage

\end{document}